\documentclass{article}
\usepackage[utf8]{inputenc}
\usepackage[a4paper,margin=1in]{geometry}
\usepackage{graphicx}
\usepackage{xcolor}
\usepackage{tabularx}
\usepackage{xurl}
\usepackage{enumitem}
\usepackage[colorlinks=true,urlcolor=black]{hyperref}
\usepackage[capitalize]{cleveref}
\usepackage{fancyhdr}

\def\BibTeX{{\rm B\kern-.05em{\sc i\kern-.025em b}\kern-.08em
    T\kern-.1667em\lower.7ex\hbox{E}\kern-.125emX}}
\vspace{-8ex}
\date{}
\begin{document}
\title{Proposals for Resolving Consenting Issues with Signals and User-side Dialogues{
    \footnotesize\thanks{This work has been funded by Irish Research Council  Government of Ireland Postdoctoral Fellowship Grant\#GOIPD/2020/790. The ADAPT SFI Centre for Digital Media Technology is funded by Science Foundation Ireland through the SFI Research Centres Programme and is co-funded under the European Regional Development Fund (ERDF) through Grant \#13/RC/2106\_P2. For the purpose of Open Access the author has applied a CC BY public copyright licence to any Author Accepted Manuscript version arising from this submission.}}}

\author{Harshvardhan J. Pandit
\\
ADAPT Centre, School of Computer Science \& Statistics \\ Trinity College Dublin\\
Dublin, Ireland \\
pandith@tcd.ie}
\pagenumbering{arabic}
\maketitle

\begin{abstract}
Consent dialogues are a source of annoyance, malicious intent, dark patterns, illegal practices and a plethora of other issues. This work presents known problems based on GDPR requirements grouped into two categories: (i) UI/UX for consenting; and (ii) power imbalance in expressing consent. To resolve this, it presents two proposals: First, the use of automation through privacy signals to better govern consenting processes and to reduce `consent-fatigue'. Second,  as generation of consent dialogues on the user side and its practicalities for both websites as well as users and agents (e.g. web browsers). Both proposals are discussed in terms of possibilities for implementation and suitability for stakeholders. The article concludes with a discussion on the difficulties in achieving such solutions owing to the conflicts of interest between `web-enablers' and `web-consumers', and the necessity for the EU to take a direct stance in addressing these in their future laws.
\end{abstract}

\section{Introduction}
\subsection{Motivation}
The mechanisms surrounding consenting on the web are today rife with problematic issues (for an overview, refer to \cref{sec:issues}) that threaten
its use as a means for organisations to source personal data, and for people to exercise their agency and rights - in compliance with privacy and data protection regulations.
The requirements established by the General Data Protection Regulation (GDPR) provide a legal framework
through which authorities can take action and dissuade their prevalence.

However, in practice, such enforcement has not been effective in terms of mitigating such issues despite the requirements of GDPR being known for 5 years (2016-2022).
This is evident from the existing prevalence of such issues, as well as the actions of consumer protection organisations and NGOs in calling for better enforcement of the GDPR. Notable examples include: The European Parliament's resolution \cite{EU_Parliament_GDPR_report_2021} on implementation of the GDPR; The Irish Council of Civil Liberties suing \cite{ICCLSuesDPC2022} the Irish Data Protection Commission with the claim of failed duties in enforcing GDPR for Real Time Bidding (RTB) as a `data breach'; and None of Your Business (NOYB) filing 422 formal complaints regarding consenting dialogues \cite{NoybFiles422a}.

This has resulted in concerns regarding the performance and effectiveness of Data Protection Authorities (DPAs), and has resulted into an official enquiry by the EU Ombudsman \cite{EuropeanCommissionFailure} to look into the EU Commission's handling of GDPR enforcement. At the same time, DPAs have (repeatedly) indicated the lack of available expertise and funding required to be effective in their duties, which has had an effect on the ability to effectively address the large amount of complaints they receive \cite{EuropeEnforcementParalysis2021a}. 

To date, typical actions undertaken by DPAs have included specific investigations into problematic areas of technologies, such as facial recognition, and to issue guidelines that call upon organisations to effectively ensure their own compliance in the hopes of self-regulation fixing existing known problems.
The few high-profile decisive cases, such as CNIL's fine to Google \cite{DeliberationSAN2021023December2022} and Facebook \cite{DeliberationSAN2021024December2022} regarding their consenting practices has failed to have any domino effect or reactive changes\footnote{To verify: as of MAR'22 Google's and Facebook's websites from EU still present a non-GDPR-compliant consent dialogue}.

GDPR allows for easing some of the investigation and compliance efforts through creation of codes of conduct and certifications. However, to date, no such mechanism exists that applies itself to remedy or prevent issues regarding consent. While IAB Europe has applied to utilise its Transparency \& Control Framework (TCF) framework as a code of conduct, the recent (FEB'22) decision by the Belgian DPA concerning the legality of TCF \cite{BEDPARestore} and its use by the advertisement industry reaffirms important issues about the legitimacy of RTB mechanisms in the first place \cite{vealeAdtechRealTimeBidding2021,vealeImpossibleAsksCan2022}. It also raises the question on how individuals and society at large do not have a medium to participate in such processes to outlay a balance between rights and requirements without getting into the legal quagmires of being controllers themselves \cite{edwardsDataSubjectsData2019}.

While all the above should be used in demanding greater effectiveness of the law through its enforcement agents (courts and DPAs), this article argues that such effectiveness should not rely solely on the capabilities of the organisations to self-regulate, but should instead \textbf{motivate socio-technical solutions that mitigate the root causes of known prolonged issues}.
This would require a change within the EU's approach towards development and utilisation standards in connection with legal compliance, where the onus is always on assisting organisations with their legal compliance tasks. While the EU has promoted standards developments in a more globally cohesive manner, such as through co-operation agreement between CEN/CENELEC and ISO\footnote{\url{https://www.cencenelec.eu/about-cen/cen-and-iso-cooperation/}}, it has refused to take any measures on the technologies and standards that underline the use of internet or its standardisation processes\footnote{The Article 29 Working Party strongly supported enforcement of DNT, see \textit{Opinion 01/2017 on
the Proposed Regulation for the ePrivacy Regulation (2002/58/EC)}. However, there has been no further development for its enforcement, and the attempt at standardising DNT failed.} (e.g. IETF, W3C). 
As a result - the capabilities, values, and features developed within such technologies do not assist with any of the requirements established by the GDPR to increase the expected level of privacy and data protection.

For example, consider the prevalence of cookie banners to fulfil requirements of the ePrivacy Directive (ePD, 2002) to provide information and management of cookies to users, and which are known to have widespread issues \cite{fouadComplianceCookiePurposes2020,matteCookieBannersRespect2020,santosAreCookieBanners2020a,soeCircumventionDesignDark2020}. 
In the 20 years since ePD was adopted, the underlying technological mechanisms in terms of how cookies can be set by websites and managed by users within their browsers have not changed in significance other than browsers attempting to prevent their usage in tracking and surveillance. To date, there is no cookie categorisation specification through which websites can clearly indicate what the cookies are necessary for, nor are there any cookie management interfaces for users within the web browsers. 
Now with GDPR, cookie banners have been supplemented with consenting interfaces that also show similar issues \cite{utzInformedConsentStudying2019,fouadComplianceCookiePurposes2020,grayDarkPatternsLegal2020,machuletzMultiplePurposesMultiple2020,matteCookieBannersRespect2020,mattePurposesIABEurope2020,nouwensDarkPatternsGDPR2020,santosAreCookieBanners2020a,soeCircumventionDesignDark2020,santosConsentManagementPlatforms2021}, and demonstrate the same pattern of no fundamental developments being undertaken to remedy them at a technological level. 

To sum all of the above, this article raises the question, ``\textbf{\textit{What  significant developments are needed to mitigate the widespread issues or malpractices associated with the application and use of consenting interfaces?}}''

\subsection{Existing Approaches}
Researchers have proposed several interesting approaches regarding how users can express, communicate, manage, control, and enforce their privacy and decisions\footnote{For example, look for personal information management system (PIMS), decision support systems, personal agents.}. However, this article focuses on concrete proposals used or proposed by stakeholder organisations so as to limit the discussion on feasible applications for real-world issues. 
The focus here is on user-side solutions as alternatives to the industry-specified standards, such as the TCF and its exclusive use by Consent Management Platforms (CMPs), which can be argued to be the cause of such issues itself \cite{vealeAdtechRealTimeBidding2021,vealeImpossibleAsksCan2022}.
This is to counter the \textit{status-quo} where industry members band together to create a solution that only they can participate within, and which is then forced upon users under the guise of a compliant framework.

Existing attempts at creating standardised approaches, such as Do Not Track\footnote{\url{https://www.w3.org/TR/tracking-dnt/}} (DNT) and Platform for Privacy Preferences\footnote{\url{https://www.w3.org/TR/P3P11/}} (P3P) are either obsolete or abandoned, with new proposals including Global Privacy Control\footnote{\url{https://globalprivacycontrol.github.io/gpc-spec/}} (GPC) and the Advanced Data Protection Control\footnote{\url{https://www.dataprotectioncontrol.org/spec/}} (ADPC) being candidates for investigation. 

GPC has been developed to be applied within California's Consumer Protection Act (CCPA) \cite{CCPA}, and is a unary (single value) signal that expresses `do not sell' as per the CCPA's enforceable obligation of providing an option to prohibit organisations from `selling' data to other third parties. GPC has been adopted by numerous high-profile websites such as the New York Times and is implemented by web browsers such as Brave and Firefox. 

ADPC is a specification for expression and communication of information regarding purposes and their use in giving or withdrawing consent and exercising the right to object. Developed by None of Your Business (NOYB) as part of the RESPECTeD project\footnote{\url{https://www.respected.eu/}}, ADPC currently does not provide the necessary information for how specific purposes should be expressed in an interoperable manner and the governance processes that should be followed by each actor i.e. websites, user-agents, and the users.

While both GPC and ADPC claim to be actionable under the GDPR and ePrivacy Directive as automated signals (i.e. GDPR Art.7 for withdrawing consent and Art.21 for right to object), there are important issues that need clarification for their effectiveness.
For GPC, the difference between its CCPA-influenced terminology and that of GDPR (e.g. sell, third-party) prevents a clear interpretation for how it applies under GDPR\footnote{\url{https://harshp.com/research/blog/gpc-gdpr-can-it-work}}.
For ADPC, a lack of implementation details prevents its realisation and possibly effective adoption, e.g. a website implementing ADPC requires certainty that the user-agent understands its purposes, which necessitates some interoperability, which ADPC does not provide. 
As the DPAs or other data protection bodies have not commented on authoritative interpretations of these proposals, it is an unresolved question as to the effectiveness of these in mitigating issues regarding consenting online.
\begin{table*}[ht]
    \centering
    \caption{Categorisation of issues related to 'consent' in SotA}
    \label{table:issues}
    \begin{tabularx}{\textwidth}{|p{2cm}|X|X|X|}
    \hline
        GDPR Clause & HCI & IT & Law \\ \hline
        freely given (R43) & nudging \cite{utzInformedConsentStudying2019,deConsentTargetedAdvertising2020,grayDarkPatternsLegal2020,humanHumancentricPerspectiveDigital2020,machuletzMultiplePurposesMultiple2020,nouwensDarkPatternsGDPR2020,santosAreCookieBanners2020a},
        consent wall \cite{grayDarkPatternsLegal2020,santosAreCookieBanners2020a,utzInformedConsentStudying2019,soeCircumventionDesignDark2020}, cannot refuse \cite{matteCookieBannersRespect2020,santosAreCookieBanners2020a},preselected options \cite{matteCookieBannersRespect2020,nouwensDarkPatternsGDPR2020,santosAreCookieBanners2020a}, nagging \cite{soeCircumventionDesignDark2020} &  & tracking wall \cite{zuiderveenborgesiusTrackingWallsTakeItorLeaveIt2017,gilgonzalezUnderstandingLegalProvisions2019,nouwensDarkPatternsGDPR2020,santosAreCookieBanners2020a} \\ \hline
        specific (R43) & wording \cite{utzInformedConsentStudying2019,deConsentTargetedAdvertising2020,machuletzMultiplePurposesMultiple2020} & cookie purposes \cite{fouadComplianceCookiePurposes2020}, ignore preferences \cite{matteCookieBannersRespect2020} & wording \cite{nouwensDarkPatternsGDPR2020,mattePurposesIABEurope2020,matteCookieBannersRespect2020,santosAreCookieBanners2020a} \\ \hline
        informed (R43) & framing \cite{grayDarkPatternsLegal2020,machuletzMultiplePurposesMultiple2020,santosAreCookieBanners2020a,utzInformedConsentStudying2019,santosConsentManagementPlatforms2021,soeCircumventionDesignDark2020}, granularity \cite{nouwensDarkPatternsGDPR2020,santosAreCookieBanners2020a}, layering \cite{nouwensDarkPatternsGDPR2020} & third party \cite{deConsentTargetedAdvertising2020}, cookie syncing \cite{urbanUnwantedSharingEconomy2018}, cookie purposes \cite{fouadComplianceCookiePurposes2020} & information required \cite{santosAreCookieBanners2020a} \\ \hline
        unambigious (R43) & \cite{utzInformedConsentStudying2019,habibEmpiricalAnalysisData2019,deConsentTargetedAdvertising2020}, action or expression \cite{santosAreCookieBanners2020a} & assume consent without action \cite{matteCookieBannersRespect2020,santosAreCookieBanners2020a}, assume given regardless of choice \cite{matteCookieBannersRespect2020,santosAreCookieBanners2020a}, transmission of signal \cite{santosAreCookieBanners2020a} &  \\ \hline
        information provision (A13,A14) & broken links \cite{habibEmpiricalAnalysisData2019}, missing information \cite{habibEmpiricalAnalysisData2019}, excessive interactions \cite{habibEmpiricalAnalysisData2019}, covertness \cite{humanHumancentricPerspectiveDigital2020} & cookie syncing \cite{urbanUnwantedSharingEconomy2018} &  \\ \hline
        legal basis &  & incorrect for implementation \cite{mattePurposesIABEurope2020} & unclear or incorrect \cite{costelloImpactsAdTechPrivacy2020,mattePurposesIABEurope2020}, applicability of consent \cite{mattePurposesIABEurope2020} \\ \hline
        withdrawal (A7) & number of actions \cite{nouwensDarkPatternsGDPR2020} & cookie \cite{santosAreCookieBanners2020a}, communication to third parties \cite{santosAreCookieBanners2020a} &  \\ \hline
        explicit consent (A9) &  &  & consent should be explicit \cite{costelloImpactsAdTechPrivacy2020,deConsentTargetedAdvertising2020} \\ \hline
    \end{tabularx}
\end{table*}

\subsection{Research Goals}
Rather than expressing an entirely new proposal, this articles takes the view that harmonised developments over existing proposals are better to consolidate stakeholder support. For this, it aims to resolve known issues and improve current consenting mechanisms by proposing novel use of existing information and communication protocols, namely HTML and ADPC respectively. Through these, it proposes a radical alternative to the current status quo of website-controlled consent interfaces where implementation of user-side consenting interfaces is argued to be useful, practical, and feasible. 
The rest of the article is structured in terms of the following research goals:
\begin{enumerate}[leftmargin=6mm]
    \item To identify, categorise, and analyse existing issues regarding consenting on the web 
    \item To address identified issues using ADPC by proposing extensions for:
    \begin{enumerate}
        \item Communication of information and decisions
        \item Establishing a shared interoperable vocabulary
        \item Generation of consent dialogues on user-side
        \item Annotating dialogues with semantic markup
    \end{enumerate}
    \item To discuss practicality of proposed work through legal, industry, and socio-technical perspectives
\end{enumerate}

\section{Categorisation of Issues}\label{sec:issues}
`Issue' in this context refers to a problematic aspect that prevents the effective or intended mechanism of consenting from realising i.e. preventing the individual from understanding and controlling the existence and extent of personal data processing associated with them.
For identification of issues, the state of the art was generated through a literature search (in Q2-Q4 2021) and reviewed using the following methods:
\begin{itemize}[leftmargin=5mm]
    \item Searching for related keywords: consent, CMP, GDPR, cookie, banner, dialogue, dark patterns, compliance
    \item Publications at relevant venues: privacy, security, data protection, law, HCI,  networking
    \item Authoritative sources: DPA (including EDPS and EDPB) guidelines, DPA decisions and filed complaints (where made available), court proceedings
    \item Secondary and Tertiary identifications through citations from collected literature
\end{itemize}

The collected references were filtered for relevance based on directly addressing issues associated with GDPR's consenting requirements, and were discarded otherwise (e.g. where dealing with ePrivacy Directive or CCPA or other consenting requirements).
These selected references were then categorised as to which GDPR requirement they addressed, with possibility to link multiple requirements for the same work. The GDPR requirements used here were: (i) Rec. 43 - freely given; (ii) Rec. 43 - specific; (iii) Rec. 43 - unambigious; (iv) Art.13, Art.14 - Information provision; (v) Art.6 - Legal Basis; (vi) Art.7 - withdrawal; and (vii) Art.9 - explicit consent. 

In this exercise, neither the collected literature nor the GDPR requirements are exhaustive or definitive, but are intended to be a preliminary collection sufficient to demonstrate the existence of categories so as to justify the proposed solutions that address them. Future work in this regards therefore involves performing an exhaustive literature search to identify issues, categories, and requirements not presented here.
The identified literature, relevance to GDPR's requirements, and their categorisation across three broad topics of HCI (e.g. interfaces, UI/UX, comprehension), IT (e.g. technical implementation), and Law (e.g. enforceability) are presented in \Cref{table:issues}. 

The following are some observations :

\noindent \textbf{Issues related to information}: terminology used in description depends on  framing of concepts, their granularity (broad vs specific), and can lead to missing or incomplete information. In this, users have no means to understand exactly what these terms mean or to provide a vocabulary they understand or are comfortable with.

\noindent \textbf{Issues related to presentation of information}: complex information, with layering used to hide problematic aspects (e.g. third-party sharing), with the placement of information and choices having a `nudging' effect. In this, users have no alternatives, must interact with given interfaces, and are faced with `consent fatigue'.

\noindent \textbf{Issues related to presentation of choices/options}: active manipulation through design choices, nudging, consent wall, disparity or imbalance between choices, preselected options, layering (hiding options). The user does not have fair choices due to being provided an interface designed for such manipulation or coercion. Even where interfaces may not be directly manipulative, them being cumbersome is enough to dissuade users from configuring options.

\noindent \textbf{Issues related to expression of choice}: no choice provided (e.g. no reject button), disparity in placement of choice (e.g. accept on 1st layer, refuse in 2nd), pre-selected options, granularity (e.g. agree all but refuse individually), layering (hiding choices in different layers), assumption (e.g. consent on scrolling or visiting), fatigue or no of actions (excessive actions e.g. to refuse all). In this, it is evident that users are aware of disparities, but gravitate towards `easy' actions for getting the dialogue out of the way. At other times, users are not aware of choices e.g. they are `hidden' in another layer or option.

\noindent \textbf{Issues related to third-parties}: hiding scale/scope of third party sharing, and insufficient information for each third-party. In this, some interfaces deliberately made it difficult to control sharing data with third-parties e.g. by using legitimate interest on consent refusal, or forcing selection over hundreds of third parties individually.

\noindent \textbf{Issues related to legal basis}: incorrect legal basis (e.g. legitimate interest instead of consent), and incorrect consent expression quality (e.g. not explicit) were predominant in practice. Users are not expected to discern the existence of such requirements or have the cognitive willingness to indulge in detection of their misuse.

From the above exercise, the proposals for solutions are broadly specified as two forms of technological innovations to address known issues:
\textbf{(1) Automating the expression of decisions} - to prevent and mitigate consent fatigue, repetitive consenting, and to ensure consistent expression of user decisions; and \textbf{(2) Providing a user-controlled method for expression of decisions} - to prevent and mitigate dark patterns, manipulative design choices, and coercion.

\section{Extending ADPC}
The ADPC specification currently defines multiple ways for obtaining information about the intended processing purposes: (1) link in HTML header to a file containing consent requests; or (2) Script that calls the consent-request API and passes the information.
The websites determine which purposes they want to use, where users (and user-agents) are required to be aware of the terms (and their variance) to enable expression of consent/object decisions. 
In absence of this, every website could implement ADPC with its own special customised purpose, thus creating information comprehension and management overload for users, especially when websites change terms or users have to manage it for every website individually.
A pragmatic solution would be using an existing common specification - the IAB's TCF\footnote{\url{https://iabeurope.eu/tcf-2-0/}}. However, the TCF has known issues \cite{santosAreCookieBanners2020a,vealeAdtechRealTimeBidding2021,vealeImpossibleAsksCan2022} and cannot be customised by websites or users. 

To rectify this, the first proposed solution consists of three incrementally radical proposals: (1) giving users control over TCF; (2) giving users and websites freedom to use any purpose - and matching them semantically; and (3) creating a globally standardised mechanism for establishing interoperability between purposes.

\subsection{Applying ADPC to IAB's TCF v2}
TCF v2 is a specification dedicated to the sharing of personal data with IAB-defined purposes. The TCF string is effectively a JSON-like policy encoded in Base64 that consists of additional info than just purposes - e.g. it also specifies parties, data categories, legal basess.
Given that ADPC only defines purposes as a single-field, and that users may want to express preferences for each of TCF's granular fields, the entirety of TCF's bit-string can be specified as a single purpose that can then be consented or objected to.
For (a simplified) example, the identifier \texttt{1A2B3C} can be the id of purpose \texttt{1A} that uses data \texttt{2B} for controller \texttt{3C}. 
Users can selectively (through their user-agent) configure ADPC to permit or reject specific parts of the TCF, such as prohibiting all uses of data \texttt{2B} or consenting only to controller \texttt{3C}.
This approach enables to utilising existing ecosystems of TCF with ADPC's signal expression, and allows existing websites and ad-mechanisms to continue operating, and provides users with more transparency and control of their choices instead of it being obfuscated in a cookie.

Since the TCF is already optimised for transmission (as binary encoding), a further optimised version can be developed by removing irrelevant fields (e.g. timestamps, metrics) for transmission through HTTP headers. The matching between a TCF-expressed request and user-expressed preference can also be made performant e.g. based on the complexity of matching, two bloom filters may suffice to simply check whether the user has expressed any preference over permitting or prohibiting an item (e.g. purpose) in the current request.

\subsection{Using DPV for Interoperability between ADPC purposes}
One of the challenges with effective use of ADPC's purposes is that is necessitates websites and user(-agents) to agree upon the vocabulary to correctly interpret the purposes. To enable some flexibility in this where websites and users can select their individual purposes and preferences, which are then semantically matched with each other to determine whether the action is permitted or prohibited. This is based on P3P's design for matching user-preferences with website requests.

For ADPC, the proposal is to use (something similar to) the Data Privacy Vocabulary (DPV) \cite{panditCreatingVocabularyData2019} - which provides hierarchical taxonomies of purposes, personal data, legal bases, and other relevant concepts. DPV's taxonomies enable one party (e.g. user) to define their preferences at an abstract or broad level (e.g. prohibit marketing), while the other party (e.g. website) can be specific for their use-case (e.g. request sending newsletters of new products). The receiving party (e.g. web browser) then performs a semantic matching that involves inferring that the request to send newsletters is prohibited by the preference to not receive any marketing, and therefore undertake consent refusal / withdrawal or express objection. This works in favour of both websites - that are required to utilise specific purposes, and users - that can set preferences at a broad level and avoid repetitive consenting decisions.

To enable the above within ADPC, requests must contain two pieces of information: the purpose as requested by the controller and the parent concept it is expanded from, e.g. \texttt{Send Newsletters :: Marketing}. In this only the top or parent concept needs to be known or interoperable between the website and user-agent, thereby reducing the requirement for shared purposes, and permitting the creation or utilisation of taxonomies like DPV that provide a broad structure of purposes.

The DPV may not be suitable for use in HTTP transmission due to a larger byte-size. Here, an approach similar to the compression utilised by TCF can be developed where a binary expression can be created by converting DPV to a `flat list' and assigning identifiers (e.g. top concept is 0, next concept is 1, etc. using the Shannong-Fano encoding\footnote{See - The Transmission of Information by Robert M. Fano (1949)}). 

Using DPV is beneficial as its taxonomy is richer than TCF, more flexible in that multiple purposes can be cleanly combined and retain individuality (i.e. multiple parent concepts), and has granularity of requests and decision management. 

\subsection{A Global Shared Registry of Purposes}
The previous two approaches assumed that the website and user-agent have a mutual understanding on which vocabulary to use.
In practicality, it may be difficult to reach agreement on a single such vocabulary given the differences in domain, sector, application, jurisdiction, and actors.
To avoid a constant struggle over which vocabulary should be used, and to avoid/prevent a single restricted vocabulary such as TCF being the only available vocabulary, a globally standardised registry of `purposes' could be established under a neutral umbrella e.g. by W3C.
The registry (and its shepherd organisation) is then responsible for maintaining the list of vocabularies conforming to some specified standards, requirements, or norms - with legal enforceability necessitating use of only registered vocabularies to avoid misuse. 

Use of registered vocabularies is similar to previous proposals - as a specification expressed in binary form similar to TCF usage to enable an optimised expression. For example, consider 4-byte words: \texttt{0110 1110 1010 0111} where the first 8 bits indicate the identifier of registered vocabulary used (\texttt{0110 1110 = 108 of 127}), with the vocabulary then specifying how the following 8 (or more) bits should be interpreted. If vocabulary 127 was TCF, then the user-agent would interpret the next 8 bits as defined by TCF's specification in the registry e.g. as each bit indicating permission for a purpose in the list (same as existing use of TCF).

The creation of vocabularies at a standardisation organisation (e.g. W3C) offers advantage in the form establishing governance procedures where authorities, NGOs, etc. can have the ability to specify additional information/requirements at the purpose level. For example, saying \texttt{x} purpose in a list \textit{must} require explicit consent, or that \texttt{y} \textit{must not} be used with legitimate interest. 

Such a global registry can also maintain `mappings' between vocabularies that permit websites to request using a vocabulary (e.g. TCF) and match it with user preferences in another (e.g. DPV). This can enable websites and user-agents to communicate in the language they deem suitable without requiring strict restrictions on communication.

\section{Generating Dialogues on User-side}
For UI/UX or HCI issues, they are directly associated with how information is presented, what and how choices are offered, and how consenting takes place (e.g. dark patterns). To rectify this, the second proposed solution consists of some or all of the consenting interface being generated and managed at the user-side by the user-agent of choice (e.g. web browser or external provider).


To better understand the implications of these solutions, a common acceptance of the terms associated with different interface elements is necessary. For the puropses of this paper: a `notice' is the entire interface which contains information; `notice elements' are parts of a notice, `controls' are actionable elements meant to be interacted with by the user (e.g. more info, select choice), and `dialogue' refers to the entire interface consisting of notice and controls. Controls can be further distinguished as: `layer controls' refers to controls associated with varying interface elements in terms of information density, granularity, presentation, etc.; `preference controls' refer to the user expressing a preference - where it is important to note that a preference is not a final decision e.g. checkboxes vs accept button; and `decision controls' that convey the decision of the user e.g. accept/reject button.

\subsection{Complete Generation}
The first approach consists of the controller providing all the necessary information in a machine-readable format (e.g. via ADPC), and receiving the user's decision (e.g. true/false). The user-agent generates the interface and presents the information and controls to the user. 

The information conveyed could be based on standardised specifications, such as those based on GDPR's Art.13 and Art.14, and using mechanisms such as the DPV \cite{panditCreatingVocabularyData2019} for interoperable communication of \textit{Purpose, Processing, Personal Data, Controller, Recipients, Legal Basis, and Tech/Org Measures}. 

Even though the user-agent generates the dialogue on user-side, the controller can still be provided with some (limited) ability to control the design and layout of individual elements (e.g. using CSS) - such as to ensure consistency with website design or to provide additional emphasis on information. Such access can be restricted to a limited set of CSS or providing specific rules on elements so as to avoid hiding elements or changing layouts for manipulation. 

This approach avoids the pitfalls related with manipulative designs and dark patterns, but interestingly also reduces the requirements for controllers to develop notice and consenting mechanisms themselves. At the same time, it also works in favour of controllers by reducing their efforts in developing notice and consent interfaces. However, the caveat here is that for controllers to respect not generating interfaces or using the user-agent's interfaces - a legal intervention is necessary.

\subsection{Providing Customisable Interfaces}
Where the earlier approach proposed completely removing the ability of controllers to create any consenting interface, this proposal relaxes this restriction to instead provide a limited ability to generate interfaces.
In this, controllers retain the ability to customise interfaces i.e. use any HTML structure, but have to use the provided framework to make requests. The term `framework' refers to some API provided by the user-agent through which the controller can generate interfaces and offer choices and controls to the user.
For example, as a standardised JavaScript call \texttt{createNotice()} that takes as parameters DOM elements containing contents of a notice - such as options for notice elements, controls, styling, etc. Like before, the final consenting decision controls are offered by the user-agent.

This approach permits more flexibility to controllers to express notices and dialogues in granular form, in a way that they determine, with their own styling. The only caveat is that they have to explicitly structure their interfaces and processes to conform with the APIs offered.
This enables user-agents to rectify or correct known issues such as disparity in consenting choices or use of layers to `hide' information. This also enables users to request consent interfaces look or behave in a certain way by controlling the use of APIs themselves e.g. by asking the user-agent to display all elements in a single screen instead of as layers.

\subsection{Only Generating Choices and Expression}
The final proposal is a vastly reduced condition where the user-agent only generates the choices and controls within an interface - which might prove to be more palatable to controllers as they retain complete control of what they want to show and how.
However, to avoid some forms of dark patterns, such as the user not having the ability to discern between choices, or being forced to interact and make a decision - the user-agent requires the controller to request consent via an API call. This is similar to the way websites can currently request access to and use microphones or cameras through the browser or on the smartphone.

An example of this could be: calling the JavaScript API \texttt{requestConsent(html\_element\_id)} where the HTML element ID refers to the ID of DOM element containing the notice and choice elements.
As before, the final decision elements are generated user-side and are not available for manipulation by the controller.

While this provides controller the ability to control how they want to make a request (and use dark patterns), the final decision making capability is specified by the user. A more granular iteration of this option is the requirement to make controllers also use the API calls to specify other controls (e.g. checkboxes) so as to prohibit manipulation of any form in making choices.

\subsection{Use of ADPC, and Quality of Consent}
In all options, ADPC (as defined) takes on a more passive role, where it can act as the bearer of information required to generate a dialogue, or for communicating the decisions made by the user. However, the proposals presented in this paper also work towards enabling consenting interfaces, such as through the use of standardised and interoperable vocabularies that can be useful in generated consenting interfaces programmatically.

This opens up an interesting aspect of research regarding the `quality of consent' - e.g. regular or explicit as defined by the GDPR. The user-agent could utilise the vocabularies and known norms (e.g. special categories require explicit consent) to automatically present the appropriate level of information and choices to the user. Not only is this beneficial for enforcement of the law (i.e. GDPR) as intended, it helps users in making them more aware of when sensitive decisions are needed, and is better for controllers to avoid incorrect or non-compliant interfaces.

\subsection{Semantic Markup in HTML documents}
Further use of vocabularies is that they can also be used as semantic markup within the HTML or externally as machine-readalbe policies. Existing proposals\footnote{\url{https://doi.org/10.5281/zenodo.5076603}} for this include embedding semantic markup alongside the HTML elements so that the user-agent can identify the notice elements and assist in the requesting and decision making process. This can be the source of information required for some of the proposed approaches above with the HTML existing as a fallback in case the user-agent does not support ADPC. In this, the explored approach consists of specifying `notice' using the \texttt{dialog} HTML element, customising buttons/choices by identifying them using semantic markup (e.g. button type as providing consent); and indicating action or application or information using \texttt{data-*} attributes (e.g. \texttt{data-purpose} indicating the text following the element is the purpose of processing).

A semantic markup mechanism already exists in the form of \url{schema.org} which provides metadata for websites to declare their contents to search-engines and other web-crawlers. However, despite aiming to offer rich semantic markup for websites, it lacks relevant concepts to represent important information such as privacy policies, t\&c, entities and roles (e.g. controller), dialogues and interfaces, consent, preferences, etc.
Therefore, alongside the creation of vocabularies and user-side generation of dialogues, a tangential proposal also includes the further development of HTML elements as identified earlier and accompanying semantic markup within schema.org so they can be used in HTML to identify notice elements and its contents e.g. to indicate which button is consent action.
Where some of these concepts could be deemed to not be in scope for schema.org, they could be explored to be alternately expressed using another vocabulary such as the DPV \cite{panditCreatingVocabularyData2019}.

\section{Discussion on Practicality and Feasibility in Real-World}
This section discusses important questions in terms of the practical application of proposed approaches, the history of similar approaches facing resilience from actors historically, as well the difficulties associated with making such approaches legally required and enforceable.

\noindent\textbf{Controllers:} While the proposed approaches make it more difficult for controllers to indulge in malpractices, this must be taken as a benefit in terms of lesser legal liability rather than a hindrance in terms of lesser data sourcing. Additionally, giving users more control may open new opportunities to build relationships based on trust rather than coercion. However, in practice, controllers may still largely rely on third-party tools such as CMPs and services such as TCF/RTB given the prevalent use of advertisements on websites. Therefore, effective adoption for controllers may be entirely determined by available support in third-party tooling. This creates a risk that proposed solutions might be rejected, diluted, or convoluted to not be effective by such third-party platforms.

\noindent\textbf{User-Agents:} User-agent here refers mainly to web browsers (i.e. Google Chrome, Mozilla Firefox, Apple Safari, and Microsoft Edge) and smartphone platform OS providers (Apple's iOS family and Google's Android). Of these, Google and Microsoft have a conflict of interest in developing such technologies given their reliance on data harvesting and ad-based revenues. For example, Chrome, despite being the dominant browser, is the last in terms of adopting beneficial approaches such as third-party cookie restrictions, but the first to offer counter proposals such as Federated Learning of Cohorts\footnote{\url{https://developer.chrome.com/docs/privacy-sandbox/floc/}} (FLoC). Collectively, the browsers have also been reluctant to improve the state of privacy-based approaches on the web as related to consenting - as evidenced by lack of related proposals or work in web communities. In addition, there is a culture of implementing things in-house, such as Apple's App Tracking Transparency\footnote{\url{https://developer.apple.com/documentation/apptrackingtransparency}} (ATT), instead of having a collective standardised approach.

\noindent\textbf{Users:} The proposed approaches promise several benefits to users ranging from lesser consenting decisions to more empowerment and better management of their relationships with controllers. However, for true effectiveness, the approaches have to be implemented by user-agents that the users commonly use and expect to work on their behalf. This is problematic because the users now face an unexpected hurdle in that they are at the mercy of their user-agents in terms of what options and abilities are made available to them. Given the historical basis for how user-agents (see prior para) have not provided any means to alleviate consenting issues, users are at a different end of power imbalance - a surprising outcome from the development of user-empowerment approaches.

\noindent\textbf{Data Protection Authorities (DPAs):} Authorities exist to enforce the law, which means that their powers, abilities, and perspectives are largely shaped within the confines of what the law declares and permits. However, authorities are also agencies in their own right, as evidenced from their lack of effective enforcement under the GDPR. While some of this can be attributed to lack of resources, the complexities of enforcing a new regulation, or the large difference in capabilities between DPAs and multi-national organisations, at the end of the day - DPAs have the power to validate solutions that they deem enforceable. While European DPAs have been reluctant to provide any reassuring endorsement to a particular solution, the Californian authority has been notable in being proactive regarding development and enforcement of technological solutions - most notably the GPC. Despite California being a smaller non-national jurisdiction and CCPA being vastly smaller than the more complex GDPR, it asserts how DPAs endorsing technological solutions can fast-track their adoption and effectiveness in legal enforcement.

\noindent\textbf{Legal Enforceability:} For solutions to be enforced, they are required to have a basis in law, such as a statement indicating applicability of a category of solutions. For example: GDPR's Art.21-5 specifically mentione ``data subject may exercise his or her right to object by automated means using technical specifications.''. Here, the exact nature of ``technical specification'' is not defined, and is open to interpretation by the courts and DPAs, but it can be taken to refer to any automated solution e.g. a signal such as ADPC. Despite this, the enforceability of a signal is still an open question as no DPA has confirmed what criteria must be fulfilled for a signal, as automated means, to be considered enforceable. There seems to be a reluctance to await adoption and support from the same stakeholders which cause issues - thereby resulting in an enforcement paralysis and a technological stagnation in terms of legally enforceable solutions. Though there are upcoming proposals that aim to change things, notably EU's Data Services Act\footnote{\url{https://digital-strategy.ec.europa.eu/en/policies/digital-services-act-package}} (DSA) and the ePrivacy Regulation\footnote{\url{https://digital-strategy.ec.europa.eu/en/policies/eprivacy-regulation}} (ePR) - their current drafts (as of MAR'22) do not provide any further clarity or ability for the use of proposed solutions.

\noindent\textbf{Standardisation:} For solutions to be generally adopted, it is essential for them to be standardised (whether formally or informally). However, the standardisation body itself does not necessarily act on its own agency. For example, IAB is a consortium made up of members who have a vested interest in advertising-based profit, and thereby a conflict on interest in any solution that reduces their capabilities. Similarly, W3C, while a web-focused body, comprises of members which include Google, Facebook, Apple, etc. that make up most of the web development task forces. While this can be mitigated through wider community participation, e.g. by NGOs, or by researchers - it is difficult to sustain such initiatives given the disparity between available human resources and funding.

\noindent\textbf{Way Forward:} While each of the prior topics was critical in that it presented barriers to the adoption of proposed solutions, it also opens discussion for overcoming these hurdles. The largest mechanism for adoption of proposed solutions is the ability for user-agents to express them, which necessitates development within web browsers (primarily) and smartphone platforms. This can be done through the W3C working and community groups dedicated to development of privacy-based features, with legal and societal backing to avoid reluctance by big-tech. At the same time, the lack of motivation stems from lesser legal enforcement and clarity - which needs to be addressed in upcoming proposals explicitly. For this, additional work demonstrating why laws and DPAs need to be progressive regarding technological solutions for enforcement is desperately needed. Finally, the proposed research is theoretical, and needs to be implemented in real-world conditions as prototypes to demonstrate its effectiveness and ability to function. For this, collaborations are needed to ensure adequate stakeholder representation (e.g. users, user-agents, and websites). Forums such as community groups\footnote{W3C Consent CG \url{https://www.w3.org/community/consent/}} provide a common ground for finding and exploiting such opportunities.

\section{Conclusion}
This paper presented two sets of radical proposals associated with issues regarding consenting it identified from the existing literature. The first proposal dealt with automated communication of information and decisions based on the ADPC specification, which addressed consent fatigue, user decision-making, and misuse of consenting. The second proposal consisted of generating consenting interfaces at the user-side by the user-agent through different variations of control available to the controller. It aimed to resolve issues regarding dark patterns, manipulation, coercion, and lack of control by users in consenting. 

To address how these proposals face hurdles in real-world adoption, the paper presented a discussion section that provided a critical overlook of how the different actors have been historically reluctant to adopt such solutions. In conclusion, it identified promise in the upcoming European legal proposals for technological approaches and solutions taking a more active role in effective consenting and its enforcement by all actors.

\bibliographystyle{ieeetr}
\bibliography{paper}
\end{document}